# Design of the KOSMOS oil-coupled spectrograph camera lenses

Thomas P. O'Brien[a], Mark Derwent[a], Paul Martini[a]
Gary Poczulp[b];
[a] The Ohio State University Imaging Sciences Laboratory;
[b] National Optical Astronomy Observatory

## ABSTRACT

We present the design details of oil-coupled lens groups used in the KOSMOS spectrograph camera. The oil-coupled groups use silicone rubber O-rings in a unique way to accurately center lens elements with high radial and axial stiffness while also allowing easy assembly. The O-rings robustly seal the oil within the lens gaps to prevent oil migration. The design of an expansion diaphragm to compensate for differential expansion due to temperature changes is described. The issues of lens assembly, lens gap shimming, oil filling and draining, bubble mitigation, material compatibility, mechanical inspection, and optical testing are discussed.
**Keywords:** optical coupling, oil-coupled

## 1. INTRODUCTION

The KOSMOS imaging spectrograph[1] utilizes a collimator/camera re-imaging optical system. The camera was designed with three optically coupled lens groups; two triplets and a doublet each employing Calcium Fluoride (CaF) and various optical glasses. The original design intent was to use Sylgard 184 as the optical coupling material. However, the Sylgard 184 failed repeatedly during testing at the -20C low temperature survival limit specified for the design and this material was therefore abandoned.

The optical performance of the camera was found to be relatively insensitive to the index of the optical coupling material. The decision was made to develop oil-coupled lens groups for the KOSMOS camera. The design was very tightly constrained due to the requirements delineated in Section 2.

## 2. DESIGN REQUIREMENTS

The KOSMOS oil-coupled camera design was unusually tightly constrained because it had evolved from a previous design using a solid optical couplant which held the lens elements together in the groups thus guaranteeing centration. The oil-coupled design was also required to use the existing lenses and was limited by the external camera volume available.

### 2.1 Opto-mechanical requirements

- The oil-coupled camera design must use the existing lenses without any modifications.

- The oil-coupled camera design must not exceed the available exterior camera volume. This required development of a very compact design approach.

- The design must be capable of centering lenses in triplets to ~ 25 micron precision and maintaining this centration over the full range of temperatures and gravity orientations.

### 2.2 Environmental requirements

- The design must survive without damage or leaks from -20C to +50C. The camera must operate within specifications from -10C to +30C.

- The design must include features (e.g. diaphragms) to accommodate the differential expansion of the aluminum cells, glass lenses, and oil without pressurizing the oil in the lens assembly

### 2.3 Oil requirements

- The coupling oil must transmit from 340nm to 1100nm with no substantial absorption for an oil optical path lengths of several hundred microns.

- The oil must be compatible with all materials (optical glasses, CaF, O-rings, diaphragms, aluminum, etc.) and not degrade optical throughput even in the near UV.
- The coupling oil must have a low enough viscosity to flow into inter-lens gaps as narrow as 75 microns in a few hours.
- The oil viscosity must remain low enough to reliably flow in and out of inter-lens gaps at temperatures as low as -20C.
- The oil should be soluble in non-toxic organic solvents and detergents to simplify cleaning.

## 2.4 General requirements

- The oil must be hermetically contained with absolutely no leaks when subjected to repeated cycling through the full survival temperature range
- Filling and draining oil from the inter-lens gaps must be simple and must not introduce any bubbles into the optical path.

# 3. DESIGN APPROACH

The KOSMOS lens groups are optically coupled using Cargille 1074LL (laser liquid) oil. The lenses are positioned using their cylindrical outside edges with silicone rubber O-rings. The O-rings provide optical centering, positioning, and oil sealing functions. To provide easy assembly and precision centering, the O-rings are compressed to their design compression level ("crush") after assembly.

## 3.1 Main camera barrel

The main camera barrel is a stepped aluminum tube with external features for kinematically mounting the camera in the instrument. Three oil coupled lens groups in modular aluminum cells are loaded into the barrel. The inter-group spacings are set with simple aluminum tubular shims. The groups are secured in the main barrel with an axially compliant clamp ring. The final rear element is a singlet field flattener held in a collet style cell on the outside face of the main barrel.

## 3.2 O-rings are the central design element

A central feature of this design is the utilization of circumferential O-rings to center the lenses while simultaneously sealing the oil. The benefits of this are a compact, simple design with a small number of parts. An additional advantage is that the radial stiffness of the O-ring can be "tuned" by changing the details of the O-ring gland dimensions or by using different durometer rubber. The risks of oil sealing, lens centering precision, and ease of assembly were retired with an experimental program.

## 3.3 Post-assembly compression of O-rings

In order to achieve the required high centering precision and high radial stiffness of the O-ring lens support, the O-rings must be compressed by about 15% of their cross section diameter. This compression is referred to as the O-ring "crush." This compression is much too high to allow a lens to be easily pushed through an O-ring in a gland with this much compression. The solution to this design dilemma is to compress the O-rings **after** a lens group is fully assembled with all lenses in place inside their respective centering O-rings. This "post-assembly compression" of the O-rings is achieved by using radial O-ring gland dimensions that give zero compression during assembly. After the lenses are in place, axial compression is applied to the O-rings by drawing together the two-piece lens group cell. The O-rings are slowly and uniformly compressed between the O-ring gland and the lens edge to the required compression level providing excellent centering, radial stiffness, and oil sealing.

## 3.4 Axial compression of lens elements and Kapton shims

The inter-lens spacing within a group is set with Kapton tape shims of controlled thickness. Three shims are placed on a lens 120 degrees apart prior to assembly. The Kapton tape adhesive holds the shims to the lens during the assembly process. After the lenses are loaded into the lens cells and the O-rings are compressed, the group axial clamp ring is installed and the axial preload is set. The axial clamp ring has engineered stiffness allowing a specified axial force to be accurately applied by rotating the threaded axial clamp ring by a predetermined angle.

### 3.5 Design Features for large temperature range operation

The oil-coupled lens group design must incorporate features that explicitly manage the consequences of differential thermal expansion and contraction over the full survival (-20C to +50C) and operating (-10C to +30C) temperature range. The main issues to address are the changes in the compression of the centering O-rings, changes in axial compression of the lenses, and the changes in the oil volume over the full temperature range.

At low temperatures, the aluminum lens cell parts contract radially and axially more than the optical glasses due to the larger coefficient of thermal expansion (CTE) of the aluminum. This results in increased compression in the centering O-rings. This effect was analyzed with the goal of maintaining O-ring compression at the nominal 15% value +/- 3%.

The change with temperature of the axial compression of the lens group is managed by using an axially compliant compression ring. The axial stiffness of this ring is engineered to provide the required axial clamping force ( ~ 5 times the lens group weight) at the nominal temperature while allowing changes in the axial clamping force of +/- 25% over the full temperature range.

The quantity of oil in the hermetically sealed volume of a lens group is fixed. However, as the temperature changes the lens cells, lenses, and oil all change in volume at different rates. If the hermetically sealed volume of a lens group were rigid, then changes in temperature would result in unacceptably large pressure changes in the oil. To prevent this a flexible element must be used to accommodate the temperature induced volume changes. This design uses two silicone rubber flat diaphragms that are free to move in and out as the oil volume changes. The volume change required was carefully calculated.

## 4. EXPERIMENTAL TEST PROGRAM

Several critical risk elements were identified with the KOSMOS camera design. In order to evaluate these risk areas an experimental program was executed with a Test Cell Assembly. The photo below shows the Test Cell Assembly that was built to evaluate this design approach.

### 4.1 Test Cell Assembly

The Test Cell Assembly was designed to test lens centering with O-rings, oil sealing, O-ring post compression, temperature performance, material compatibility, etc.
The Test Cell can mount a stepped aluminum cylinder "dummy lens" for measurement of lens centering and post compression or a CaF/fused silica doublet used for testing of lens spacing shims, oil filling, oil performance at low temperature, and sealing against fine ground lens edges.

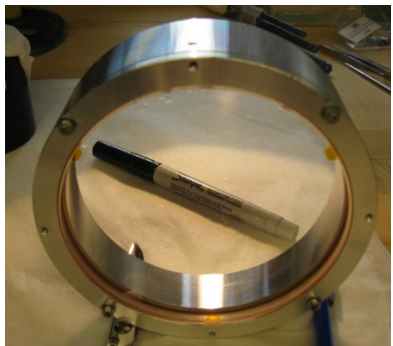
Figure 1: Test Cell Assembly with CaF/Fused Silica Doublet installed

### 4.2 Lens centering performance of post-compressed O-rings

The optical design of the KOSMOS cameras is very sensitive to lens centering. The optical centering of the lens elements should be within ~25 microns radial offset (Total Indicated Runout (TIR) = 50 microns) for a ~10% image degradation (per Zemax analysis).

A Test Cell Assembly was designed to allow very precise evaluation of the centering performance of a pair of compressed O-rings. The test article included a two piece aluminum lens cell to apply post-assembly compression on the O-ring and a cylindrical aluminum "dummy lens".

The Test Cell Assembly with dummy lens was assembled and mounted on a precision spindle and instrumented with two electronic indicator probes. The TIR differential radial run-out was measured between the lens cell and the dummy lens. The dummy lens centering accuracy was measured to be between 16 and 36 microns TIR during several trials. This corresponds to 8-18 microns of lens de-center, exceeding the design requirement by approximately a factor of two. The photo below shows the test setup.

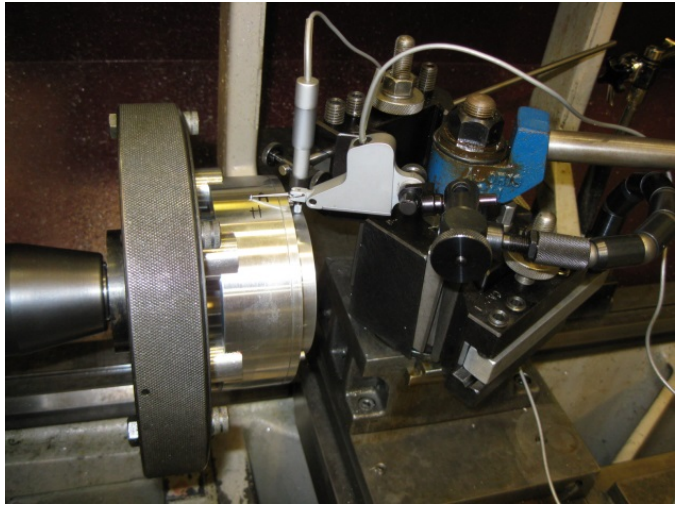

Figure 2: Test Cell Assembly with 'dummy lens" mounted on spindle with digital gauges to measure centering accuracy.

### 4.3 O-Ring Radial Stiffness

The two O-rings in the Test Cell Assembly are #246 (0.136" diameter) A70 durometer silicone rubber from Parker. Each O-ring is axially compressed after assembly by 0.5mm resulting in ~15% crush. The radial stiffness of the O-rings was measured using a force gauge to measure the applied radial force and differential indicators to measure the lens radial displacement. Several radial force values were applied at two different O-ring crush values, 15% and 7.5%. The results are shown in the graph below.

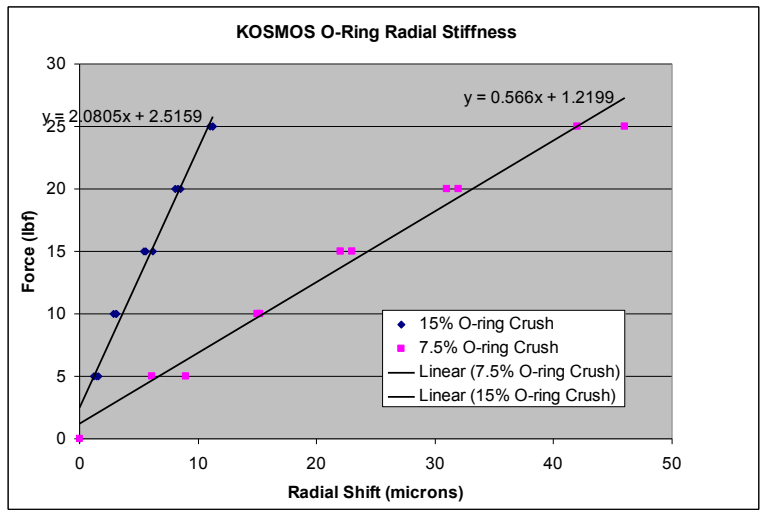

Figure 3: O-ring radial stiffness for two O-ring compression values (% crush).

At the design nominal of 15% crush the radial stiffness is ~2 lbf /micron for two O-rings working together. Since the lens triplet groups weigh about 5 lbs, they will displaced by about 2.5 microns when moved from Zenith to horizon. However, all the groups will largely move together (common mode due to gravity) so the relative de-center will be negligible and the image motion will be only a few microns.

The conclusion is that the O-ring radial stiffness at 15% crush is more than adequate to preserve optical alignment and keep image motion acceptably low.

### 4.4 O-ring Sealing and Oil Leaks

The O-rings are used to seal against the fine-ground outside edge of the test doublet lenses. The two lenses are made of CaF and fused silica. The fused silica lens has a moderately poor surface finish on the edge with patterned grinding marks. No leaks were detected and the O-ring/glass interface looks excellent when viewed at 8x magnification.

No oil leaks whatsoever occurred during low temperature testing to -20C. The hermeticity of the O-ring/ground glass seal is excellent.

### 4.5 O-ring Temperature Change Accommodation

The silicone O-rings have a rated temperature range of -54C to +218C. The flexibility of the O-rings has been qualitatively checked at -25C, there is no apparent change relative to room temperature flexibility.

The O-ring crush will change from the nominal value of 15% @ 20C to ~13.5% at +60C and ~ 16.5% at -20C. Thus the full range of temperatures has only a small effect on O-ring crush. This minimizes de-center, stiffness, and sealing problems.

### 4.6 Oil Volume Change Compensation

The volume of the oil cavity in the largest KOSMOS triplet will change by ~ +/- 0.62cc at the two survival temperature extremes. This must be accommodated with a flexible element that is referenced to atmospheric pressure. A silicone rubber diaphragm supported over a machined cavity performs this function.

A test diaphragm assembly was designed, built, characterized, and tested. Leak testing was done by pressurizing the diaphragm with air and checking for bubbles with soap solution. The test diaphragm works well, is leak free, and is easy to assemble.

Shown in the photo below are the test cavity (which includes two fill/vent holes), the 1.5mm thick silicone rubber diaphragm, and the diaphragm clamp ring with clamp screws

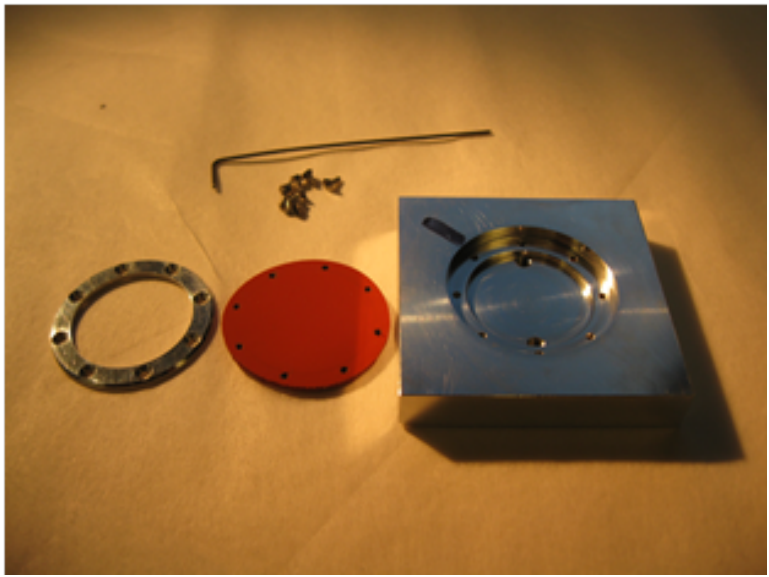

Figure 4: Test diaphragm assembly showing silicone rubber diaphragm, clamp ring, and test cavity.

The working volume of the compensator is the range of volume change over which the diaphragm develops virtually zero pressure change. This was measured with a syringe and the value is approximately 1.2cc total (+/- 0.6cc) per compensator. With two compensators per lens group the design has a margin of ~2X.

Two diaphragms of this design were added to the Test Cell Assembly. The volume compensators worked well during high and low temperature cycling with no leaks. Oil pressure was not measured.

### 4.7 Oil Filling and Bubble Management

The lens doublet and triplet cells are filled with oil prior to installation into the main camera barrel. The Cargille 1074 oil is introduced into the oil cavity from the bottom of the assembly while air is vented from the top. A 0.75mm hole into the oil cavity is backed with an M4 tapped hole for the fill fitting and fill plug. A M4 nylon tipped set screw is used as the final sealing plug after oil filling.

All features of the lens cell (O-ring grooves, volume compensators, spacers, etc) are intended to allow venting of extra oil and any entrapped air through the top vent hole during filling.

The doublet test lens with a 90 micron lens gap set by Kapton tape shims has been filled several times. The lens can be manually slowly filled in an hour with a syringe. Several "perfect fills" have been achieved with no air bubbles or pockets. During the fill process some extra oil with entrained air bubbles does vent from the top vent hole. Extra oil can be vented until no more bubbles exit the vent hole with the oil.

There was a single 0.5mm bubble near the edge of a Kapton tape shim and a second 0.5mm bubble in the exterior chamfer of the doublet. Neither of these bubbles moved or changed shape after repeated temperature cycling from room temp to -26C and from room temp to +45C. The fact that the bubble in the chamfer did not enter the lens gap is very important. Any bubbles in the chamfer are stable due to their minimum energy spherical shape and hence will not enter the lens gap. These bubbles are out of the optical path and are therefore not of concern.

### 4.8 Test Cell Assembly Temperature Testing

The fully assembled and oil filled Test Cell was taken from room temperature to -26C three times. The cool-down was achieved in a brute force manner by simply packing the assembly in ~2mm of insulation and placing the assembly in a freezer at -26C.

- No change in appearance was noted.
- Bubbles did not migrate.
- No leaks of oil were detected.

The fully assembled and oil filled Test Cell was taken from room temperature to +46C, +52C, and finally to +58C. The warm-ups were achieved by simply packing the assembly in ~2mm of insulation and placing the assembly in a pre-heated oven at the target temperature.

- No change in appearance was noted.
- Two bubbles in the lens interface chamfer did migrate circumferentially but did not enter the 90 micron oil gap
- No external leaks of oil were detected from either the O-ring/lens seals or from the diaphragms.

### 4.9 Oil/Rubber material compatibility

The RSS spectrograph on Salt experienced a reduction in UV throughput which was eventually attributed to contamination of the Cargille 1074LL optical coupling oil by reaction with urethane and Viton[2]. A test program was conducted with many material combinations. Silicone rubber O-rings were tested for oil compatibility by immersion for hundreds of hours in Cargille LL5610 & LL3421 laser liquids. There was no UV transmission drop down to 280nm. Therefore the Cargille oil with silicone rubber O-rings were selected for the KOSMOS design.

The only other potentially reactive material in contact with the oil is Kapton tape with silicone adhesive. In order to demonstrate that the Kapton tape does not have a material compatibility problem with the coupling oil, nine tape tabs were immersed in Cargille oil for three months with no peeling of tape from the lens or any degradation of the Kapton or the silicone adhesive.

The 1074LL oil is a mineral oil not a silicone based oil. It can be cleaned with organic solvents such as acetone and can also be removed with detergent and water.

### 4.10 Lens Spacing with Kapton Tape Shims

The lens gaps are set with three radial tape tabs made from 6.3mm wide, 50 micron thick Kapton tape with 38 micron thick silicone adhesive. The tape tabs are pre-assembled and burnished onto the lens surfaces with a cotton swab. One can visually confirm that there are no wrinkles on the tabs prior to assembly and oil filling. The tabs are captive and under compression after assembly.

The effective thickness of the tape was measured using two methods. The first was simply to burnish a strip of tape to a surface plate and measure its thickness with a precision probe. The measured thickness value (Kapton plus adhesive) was 85 microns. The second and more realistic measurement method involved placing three Kapton tape shims on a circular flat mirror and using the shims to space a second flat mirror. The gap produced by the three tape shims was measured as 85 microns.

## 5. FINAL DESIGN

The final design of the KOSMOS camera satisfies all the design requirements discussed above and incorporates all the design elements that were evaluated in the experimental test program. The figure below shows a cross-section view of the nine-element camera with many of the key parts labelled.

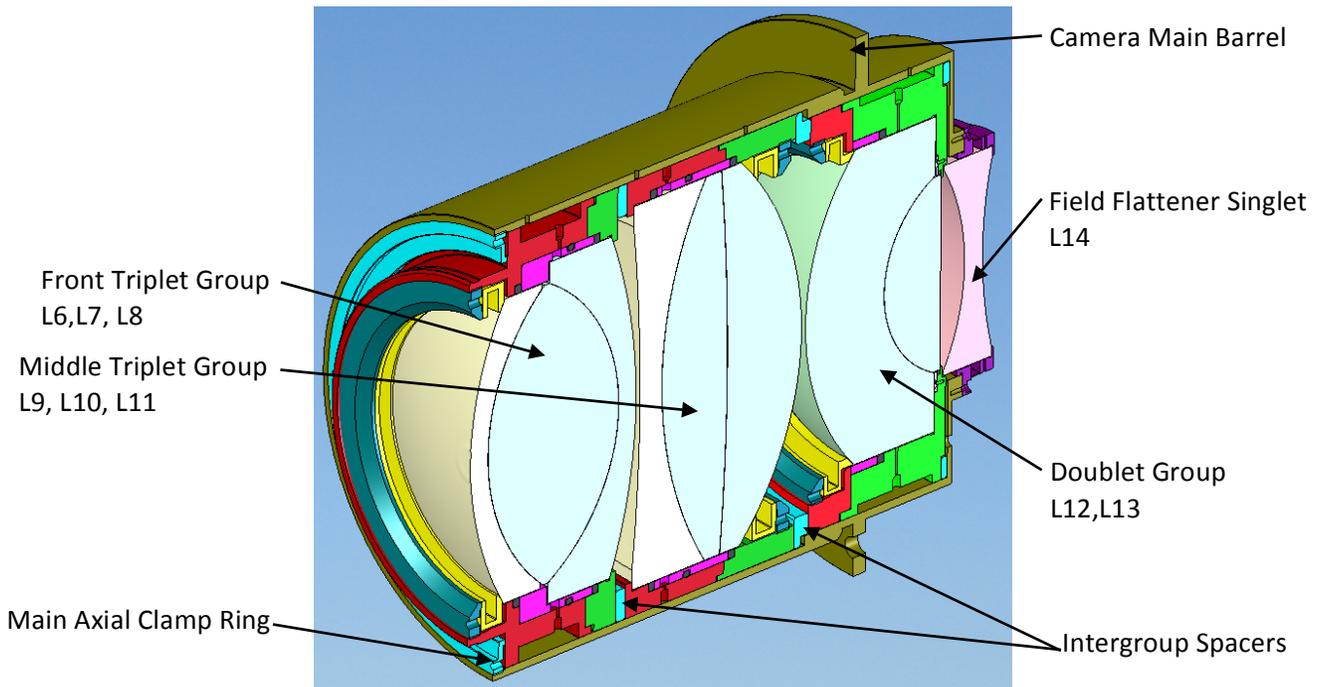

Figure 5: KOSMOS Main Camera Assembly, 2 triplets, 1 doublet, field flattener.

A detailed cross section view of the first triplet group of the KOSMOS camera is shown in the figure below. This section was taken through the volume compensating diaphragms. This triplet group is fully assembled and filled with oil prior to being loaded into the camera main barrel.

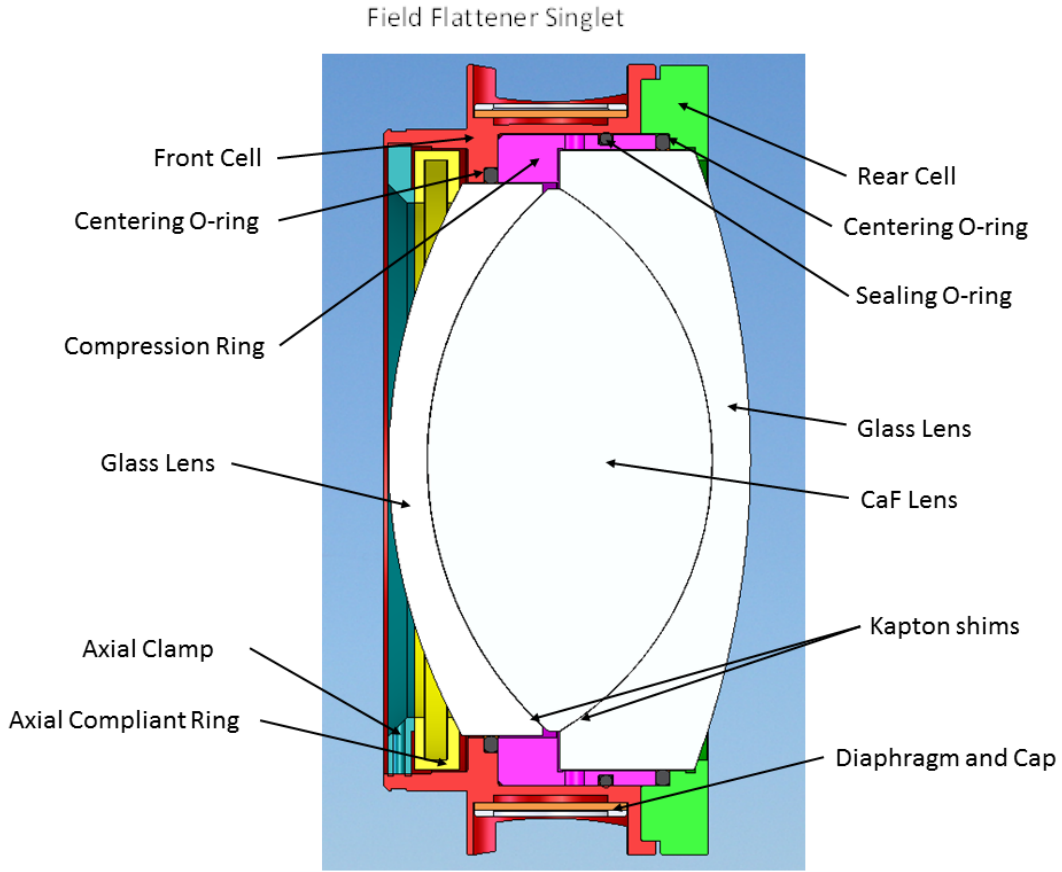

Figure 6: KOSMOS camera front triplet group assembly.

## 6. ASSEMBLY OF CAMERA

The assembly and filling with oil of the triplet and doublet groups requires a carefully planned multi-step procedure using special handling tools and fixtures. The careful placement of O-rings over the lens edges also requires careful technique. The photos and captions below illustrate the basic assembly sequence.

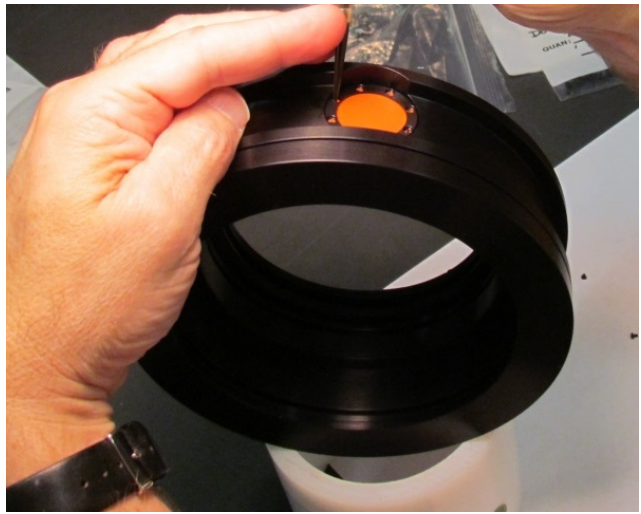

Figure 7: Diaphragms with caps are assembled to Front Cell and leak tested with pressurized air.

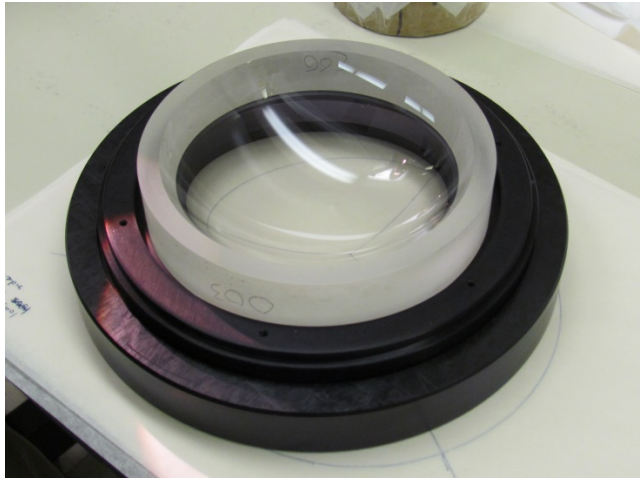

Figure 8: L8 is placed into Rear Cell.

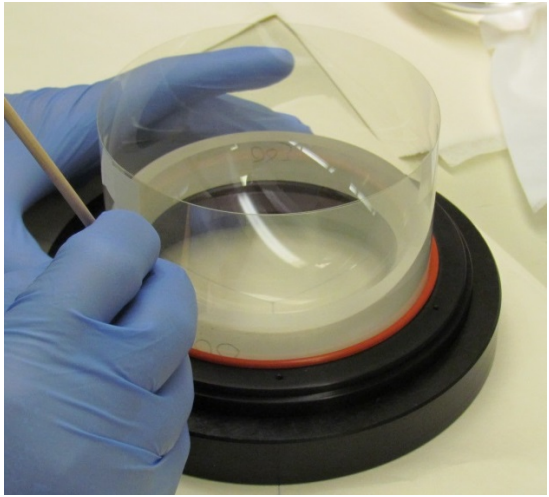

Figure 9: The centering O-ring is guided into place with a thin mylar tube which is wrapped around the L8 lens circumference. The O-ring is rolled off the tube onto the edge of the lens.

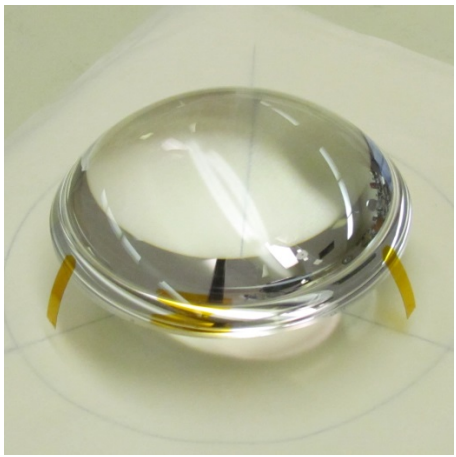

Figure 10: Kapton tape shims are placed and burnished onto both sides of L6 CaF lens prior to assembly, then trimmed. A thin film of oil is applied to the shims to prevent the Kapton shims from sticking to the mating lens.

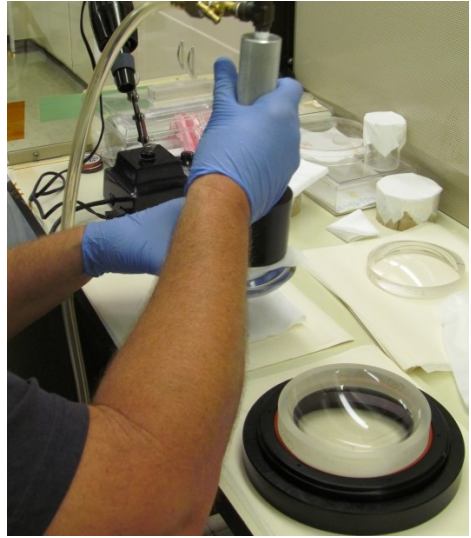

Figure 11: Lens L7 is lifted with a vacuum chuck and lowered into the concave side of L8.

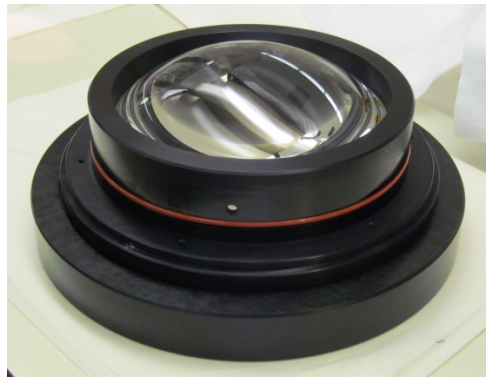

Figure 12: The Compression Ring is lowered over L7 lens. The dimensions of the cells and lenses are intended to produce zero crush on the centering O-rings during assembly.

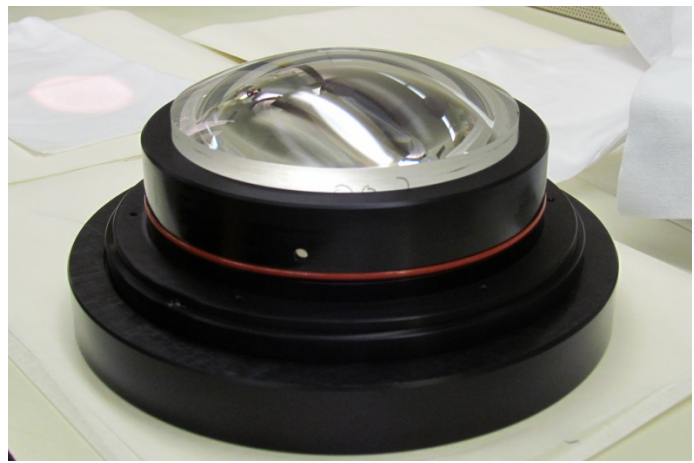

Figure 13: Install L6 onto L7 using the vacuum chuck.

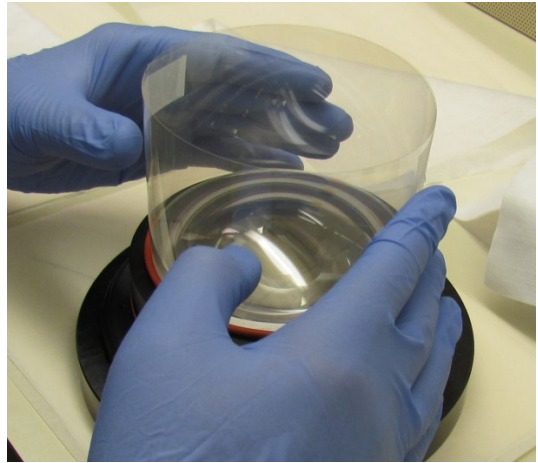

Figure 14: The L6 Centering O-ring is guided into place with a thin mylar tube which is wrapped around the lens circumference. The O-ring is rolled off the tube onto the edge of the lens.

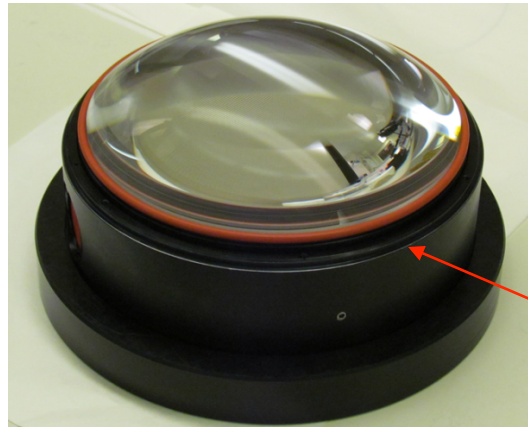

Centering O-ring

Figure 15: Front triplet showing L6 Centering O-ring on lens circumference.

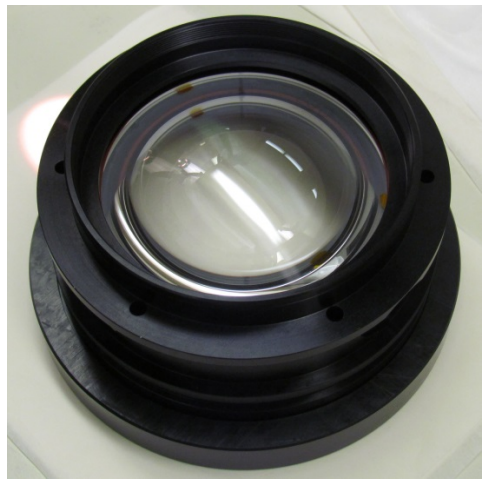

Figure 16: The Front Cell is assembled onto Rear Cell. Tightening the six axial screws in a careful slow symmetric pattern compresses the O-rings uniformly and centers the lenses very accurately and rigidly.

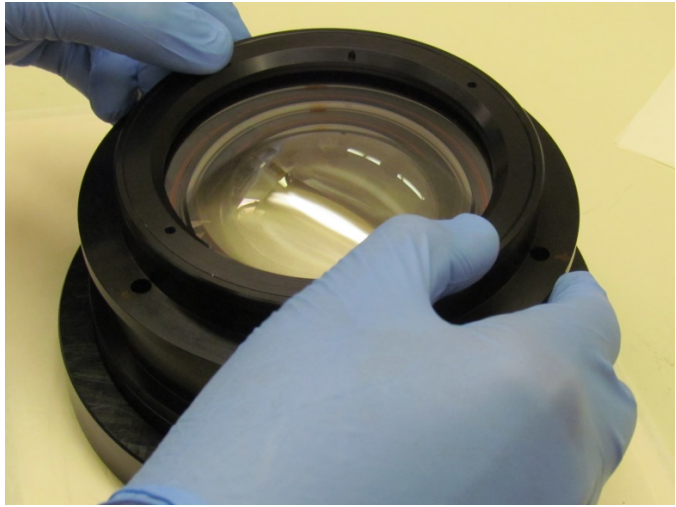

Figure 17: The Axial Compliant Ring and Axial Clamp are screwed into the Front Cell. This generates the axial preload force through the triplet, preloading the Kapton shims and ensuring that the lens gap is accurately set and maintained.

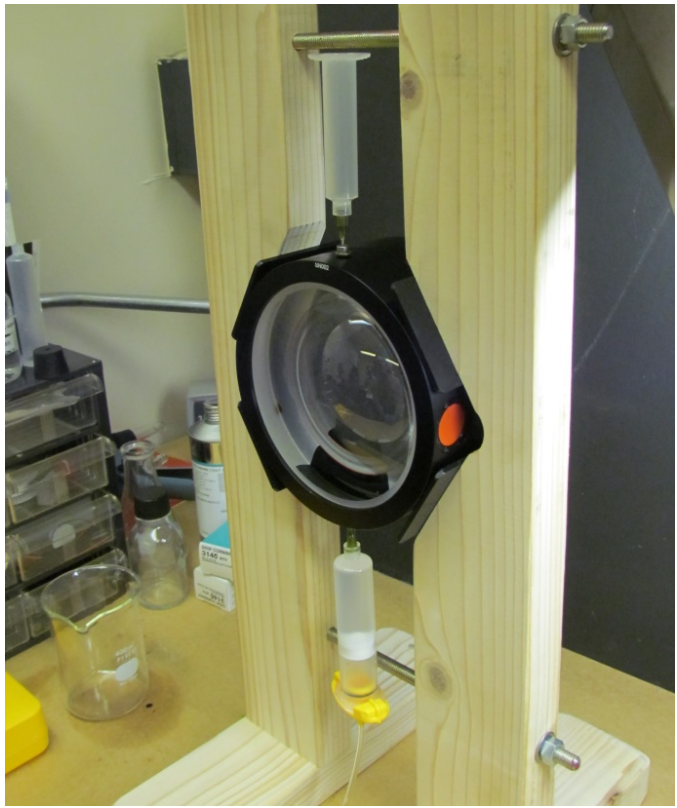

Figure 18: A triplet in the oil filling stand showing the oil filled syringe at the bottom and the overflow syringe at top. Regulated pressurized air pushes the syringe very slowly to fill the lens assembly.

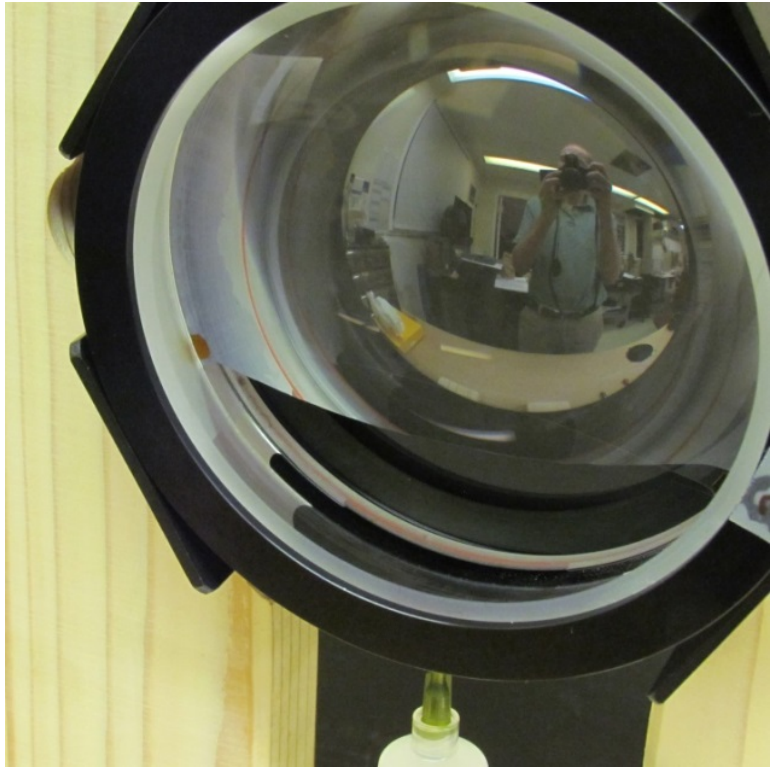

Figure19: Partially filled lens showing meniscus line of oil, about 1/3 full. Note the absence of bubbles during filling.

## 7. OPTICAL TESTING AND FINAL IMAGE QUALITY

The KOSMOS instrument was optically tested end to end in the lab. The image quality was measured at NOAO with a pinhole mask mounted in the slit wheel that was illuminated over a broad wavelength range by an external light source. The pinhole mask consists of 27 pinholes distributed on a grid that approximately sample the full extent of the telescope focal surface. The pinholes are circular apertures 5um in diameter, and thus the illuminated pinholes are effectively point sources (the focal plane scale is 6.6 arcseconds/mm). The location of each pinhole along the optical axis was adjusted to match the modest curvature (R=3.33m) of the telescope focal surface.

We obtained a camera focus sequence of eleven pinhole mask images with 25 micron steps that went through the best focus position. We measured the image size of each pinhole as a function of camera focus and used these data to determine the best image quality, best focus position, and search for any tilt between the slit wheel and detector, as well as off-axis aberrations. Several measurements were made to confirm that the image quality does not have significant wavelength dependence.

All of the pinhole images were clearly under-sampled at the best focus position (the scale is 0.29 arcseconds/pixel), and no evidence of tilt or off-axis degradation was seen. This is consistent with the ZEEMAX optical design model predictions and demonstrates compliance with the image quality requirement of 0.5 arcsecond FWHM from 400nm to 1um.

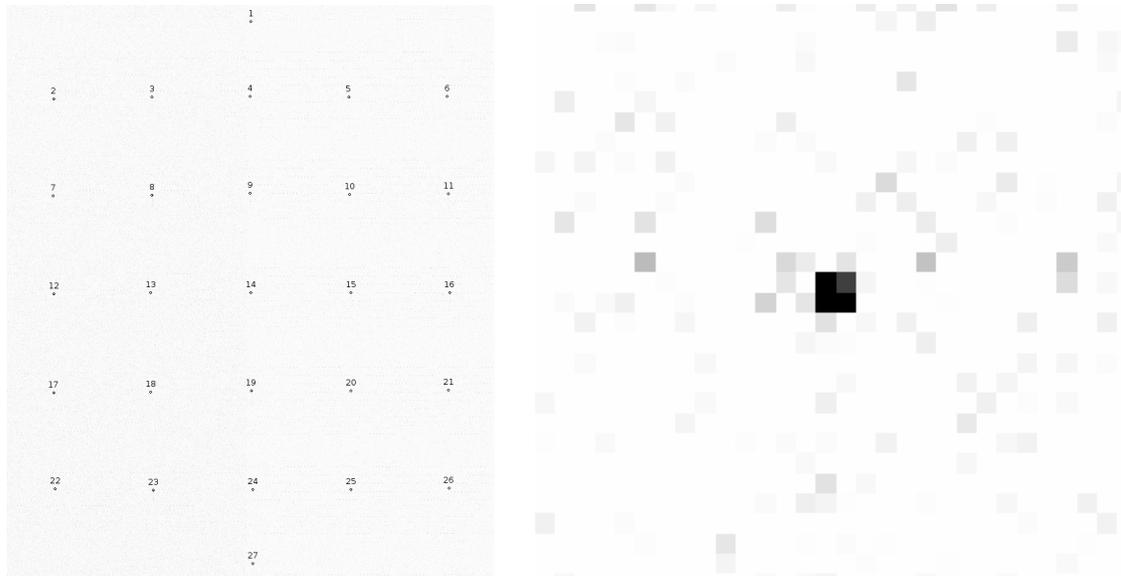

Figure 20: (Left) Image of the KOSMOS full field pinhole mask layout used in lab testing. (Right) Lab image of a single pinhole at best focus showing the under-sampled PSF has a FWHM < 2 pixels.

## REFERENCES

[1] Paul Martini, Jay Elias, et al. "KOSMOS and COSMOS: new facility instruments for the NOAO 4-m telescopes" Proc. SPIE 9147-34 (2014)
[2] Kenneth Nordsieck, Frank Nosan, J. Alan Schier, "Ultraviolet compatibility tests of lens coupling fluids used in astronomical instrumentation" Proc. SPIE 7735-82 (2010)
[3] Stephen A. Smee, James E. Gunn, et al. "The Multi-Object, Fiber-Fed Spectrographs for SDSS and the Baryon Oscillation Spectroscopic Survey" AJ, 146, 32 (2013)